\documentclass{pasj01}

\usepackage{lineno}
\Received{$\langle$reception date$\rangle$}
\Accepted{$\langle$acception date$\rangle$}
\Published{$\langle$publication date$\rangle$}
\begin{document}
\title{\LETTERLABEL
Probing Cosmic Rays with Fe K$\alpha$ Line Structures Generated by Multiple Ionization Process}
\author{Hiromichi \textsc{Okon}\altaffilmark{1}}
\author{Makoto \textsc{Imai}\altaffilmark{2}}
\author{Takaaki \textsc{Tanaka}\altaffilmark{1}}
\author{Hiroyuki \textsc{Uchida}\altaffilmark{1}}
\author{Takeshi~Go \textsc{Tsuru}\altaffilmark{1}}%
\altaffiltext{1}{Department of Physics, Kyoto University, Kitashirakawa-oiwake-cho, Sakyo-ku, Kyoto 606-8502, Japan}
\altaffiltext{2}{Department of Nuclear Engineering, Kyoto University, Katsura, Nishikyo-ku, Kyoto 615-8540, Japan}
\email{okon@cr.scphys.kyoto-u.ac.jp}

\KeyWords{atomic processes --- radiation mechanisms: non-thermal --- line: formation  --- cosmic rays --- ISM: supernova remnants}

\maketitle

\begin{abstract}
Supernova remnants (SNRs) have been regarded as major acceleration sites of Galactic cosmic rays.
Recent X-ray studies revealed neutral Fe K$\alpha$ line emission from dense gas in the vicinity of some SNRs, 
which can be best interpreted as K-shell ionization of Fe atoms in the gas by sub-relativistic particles accelerated in the SNRs. 
In this Letter, we propose a novel method of constraining the composition of particles accelerated in SNRs, which is currently unknown.  
When energetic heavy ions collide with target atoms, their strong Coulomb field can easily cause simultaneous ejection of multiple inner-shell electrons of the target. 
This results in shifts in characteristic X-ray line energies, forming distinctive spectral structures. 
Detection of such structures in the neutral Fe K$\alpha$ line strongly supports the particle ionization scenario, and furthermore provides direct evidence of heavy ions in the accelerated particles.
We construct a model for the Fe K$\alpha$ line structures by various projectile ions utilizing atomic-collision data.
\end{abstract}

%\linenumbers
\section{Introduction}
%Cosmic rays (CRs) are one of the major energy sources in the interstellar space, and hence play an essential role for the structure and evolution of our Galaxy.
Supernova remnants (SNRs) have been regarded as major acceleration sites of Galactic cosmic rays (CRs). 
Radio, X-ray, and gamma-ray observations have been providing evidence that particles are indeed accelerated in expanding shells of SNRs via the 
diffusive shock acceleration mechanism (e.g., \cite{Koyama1995,Aharonian2007,Ackermann2013}). 
However, the emission channels detected are all radiations from relativistic particles, and thus particles with lower energies 
were almost unexplored until recently. 

Suzaku data recently revealed the presence of enhanced neutral Fe~K$\alpha$~($\sim$ 6.4 keV) line emission in some SNRs, where gamma rays are detected and thus 
particle acceleration is expected to be at work (e.g., W44: \cite{Nobukawa2018}; W28: \cite{Okon2018}; G323.7$-$1.0: \cite{Saji2018}).
In most cases, the line emission spatially coincides with dense gas in the vicinity of the SNRs. 
Based on these results, the authors interpreted that the Fe line emission is due to K-shell ionization of Fe atoms in the gas by particles accelerated in the SNRs.
If the interpretation is the case, sub-relativistic particles are mainly responsible for the Fe emission line since the production cross sections of the line emission 
peak at $\sim 10~{\rm MeV}$ and $\sim 20~{\rm keV}$ for protons and electrons, respectively \citep{Dogiel2011}.

Information on accelerated sub-relativistic particles can be exploited using the neutral Fe line emission. 
\citet{Dogiel2011} proposed an idea that species of radiating particles can be distinguished based on the equivalent width of the Fe line with respect to 
a non-thermal bremsstrahlung continuum from the same population of particles. 
Using the idea, \citet{Nobukawa2018}, \citet{Nobukawa2019}, and \citet{Saji2018} claimed that protons account for the majority of the Fe line emission in some SNRs.
\citet{Makino2019} presented an analytical model in which they consider energy-dependent escape of CRs from an SNR shock into an interacting cloud, and calculated emission 
spectra from the cloud, i.e., Fe line emission from sub-relativistic protons and $\pi^0$-decay emission from relativistic protons. 
Applying their model to W44 and W28, they successfully reproduced both the observed Fe line emission intensity and the gamma-ray spectra. 

Fine structures of the Fe~K$\alpha$ line, if detected, can make another step forward in the study of sub-relativistic particles accelerated in astrophysical objects.
\citet{Tatischeff2012} pointed out that charge exchange between fast heavy ions and ambient neutral atoms can produce broad line structures accompanying with narrower K-shell lines. 
Detection of such structures would allow us to prove the presence of heavy ions in accelerated particles and to constrain their composition. 
Apart from charge exchange, multiple ionization, which is the topic of this Letter, is another potentially 
important process caused by accelerated heavy ions, and thus can be a powerful diagnostic tool to 
probe accelerated particles. 

In the present work, we qualitatively estimate the effect of the multiple ionization process on the Fe~K$\alpha$ line emission. 
We calculate a model to predict line structures due to multiple ionization by using the knowledge from beam experiments.
Based on our model, we show Fe~K$\alpha$ line structures expected to be detected in SNRs by high resolution X-ray spectroscopy 
with XRISM.  

\section{K$\alpha$ line structures by impacts of mono-energetic ions}
\subsection{K$\alpha$L$^i$ peaks due to multiple ionization process}
\begin{figure}
\begin{center}
\includegraphics[width=8cm]{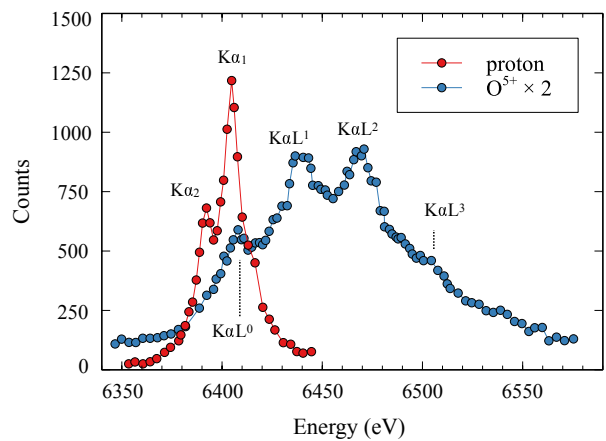} 
\end{center}
\caption{
Fe K$\alpha$ spectra produced by impacts of protons (red), and O$^{5+}$ ions (blue) \citep{Burch1971}.
For a display purpose, the O$^{5+}$ spectrum is scaled by a factor of two. 
}
\label{fig:Fig1}
\end{figure}

Ionization process of target atoms by projectile ions have been widely studied through both experimental and theoretical approaches.
With studies by high-resolution crystal spectrometers, authors such as \citet{Burch1971} and \citet{Kauffman1973} found differences of line structures between proton- and ion-produced K$\alpha$ spectra.
Figure~\ref{fig:Fig1} shows spectra of Fe~K$\alpha$ line structures originally presented by \citet{Burch1971}. 
The proton-produced spectrum has only the neutral Fe~K$\alpha_1$ and Fe~K$\alpha_2$ lines due to the K-L$_1$ and K-L$_2$ transitions, respectively. 
The O$^{5+}$-produced spectrum, on the other hand, has broad line-like structures in the 6350--6550~eV band. 
When projectile particles are heavy ions, their strong Coulomb field can easily cause simultaneous ejection of multiple inner-shell electrons of targets, resulting in the significant characteristic structures.

The multiple ionization structures consist of peaks called K$\alpha$L$^i$ ($i$ = 0--7), each of which is superposition of the K-L$_1$ and the K-L$_2$ transition lines from ionized Fe atoms 
with $i$ L-shell vacancies.
The center energies of the K$\alpha$L$^i$ peaks are higher by $\sim i \times 20$--$30~{\rm eV}$ than those of 
the Fe K$\alpha_1$ and K$\alpha_2$ lines.
This is because the K-L$_1$ (K-L$_2$) transition lines shift upwards by $\sim 20$--$30~{\rm eV}$ per L-shell electron ionization ($=\Delta E_{\rm L}$) (e.g., \cite{Wang2012}).
The outer shell (M, N, ...) ionization is mainly manifested in broadenings of the K$\alpha$L$^i$ peaks because 
the energy shift $\Delta E_{\rm M}$ (and $\Delta E_{\rm N}$, ...) of the transition lines due to additional M-shell and outer shell vacancies %for medium and large $Z$ atoms
are sufficiently smaller than $\Delta E_{\rm L}$, and is equal to or smaller than the natural widths of the transition lines.

\subsection{Modeling of K$\alpha$L$^i$ peaks} \label{subsection:2-2}

The K$\alpha$L$^i$ peaks are the sums of the K$\alpha_1$L$^i$ and K$\alpha_2$L$^i$ sub-peaks, which are superpositions of the K-L$_1$ and K-L$_2$ transition lines, respectively.
The K$\alpha_1$L$^i$ (K$\alpha_2$L$^i$) sub-peaks are expressed as superpositions of Lorentzians corresponding to transition lines.
According to \citet{Horvat2006} and \citet{Horvat2009}, K$\alpha$L$^i$ spectra obtained in ion beam experiments can phenomenologically be reproduced by the sums of the K$\alpha_1$L$^i$ and K$\alpha_2$L$^i$ peaks described by Voigt functions, or a convolution of a Lorentzian  and a Gaussian.
Here, the Gaussian reflects broadening of the sub-peaks due to simultaneous ionization of electrons in outer shells.
Taking the same approach, we modeled the K$\alpha_1$L$^i$ and K$\alpha_2$L$^i$ sub-peaks with the Voigt function.

We calculate the center energy of the sub-peaks with the scaling law given by \citet{Horvat2006}.
They expressed the energy shift $\Delta E^i {\rm (eV)}$ of each K$\alpha_1$L$^i$ (K$\alpha_2$L$^i$) sub-peak with respect to the neutral K$\alpha_1$ (K$\alpha_2$) line emission based on various data obtained by ion beam experiments.
The energy shift $\Delta E^i$ can be described as
\begin{equation}
\Delta E^i = p_L^X\, [(i-1)(Z_2-a) + b] +  i(c + d\,Z_2 + e\,i + f\,i\,Z_2)~{\rm eV}, \label{eq:delta_E}
\end{equation}
where $a=9.11$, $b =14.3$, $c=-11.64$, $d=1.493$, $e=0.755$, $f=-0.0112$, and $Z_2$ is the atomic number of the target.
The parameter $p_L^X$ is the ionization probability per L-shell electron in K-shell ionization.
We calculate the ionization probability $p_L^X$ with the model by \citet{Horvat2006}, 
\begin{equation}
p_L^X = a^{\prime}/[1+(b^{\prime}/X_2)^{c^{\prime}}], \label{eq:plx}
\end{equation}
where $a^{\prime}=0.537$, $b^{\prime}=2.11$, and $c^{\prime}=2.02$.
The parameter $X_2$ is a universal function derived by \citet{Sulik1987}, and is described as 
\begin{equation}
X_2 = 4V[G(V)]^{1/2}Z_1/(2v_1),  \label{eq:xn}
\end{equation}
where $v_1$, $Z_1$, $V$, and $G(V)$ are the projectile velocity in atomic units, where is normalized to the Bohr velocity $2.18\times10^6~{\rm m/s}$, the atomic number of the projectile, the ratio of $v_1$ to the average velocity of electrons in target atoms, and the Gryzinski Geometrical factor, respectively.
We employ an analytical formula in \citet{McGuire1973} for $G(V)$ .

We then calculate the widths of the sub-peaks.
The scaling law by \citet{Horvat2006} approximates the widths of the Gaussian of the Voigt function as 
\begin{equation}
\sigma^i = a\,i\,(b-i)(Z_2-c)~ {\rm eV}, \label{eq:sigma}
\end{equation}
where $a=0.0246$, $b=9.86$, and $c=10.40$.
Following \citet{Horvat2006}, we fix the line width of the Lorentzian component to the natural widths of the transition lines obtained by \citet{Campbell2001}.

The above scaling laws by \citet{Horvat2006} are derived from experimental data using a variety of solid targets with $Z_2 = 17$--$32$ and projectile ions with $Z_1 = 6$--$83$ at $2.5$--$25~{\rm MeV}/{\rm amu}$.
The data cover impacts by various ions with $\sim 10~{\rm MeV}/{\rm amu}$, where the cross sections for K-shell ionization of Fe atoms ($Z_2$) peak. 
%The data cover impacts by various ions with $\sim 10~{\rm MeV}/{\rm amu}$, where the cross sections for K-shell ionization of Fe atoms ($Z_2$) peak. 
%We applied the scaling laws to the calculation of the Fe line structures generated by impacts of various ions at $0.5$--$1000~{\rm MeV}/{\rm amu}$.
We applied the scaling laws to the calculation of the Fe line structures generated
by impacts of various ions in $0.5$--$1000~{\rm MeV}/{\rm amu}$ region, where collision nature is well-described by the perturbation theory and the scaling laws by \citet{Horvat2006} can be considered valid.
%Although the experimental data do not cover the energy ranges of $< 2.5~{\rm MeV}$ and  $> 25~{\rm MeV}$, ions with such energies have minor contributions 
%to the Fe line. We also confirmed that the results are qualitatively reasonable even in the energy ranges. 

%%%
%while the data do not cover impacts by fast ($\geq25~{\rm MeV}/{\rm amu}$) and slow ($\leq2.5~{\rm MeV}/{\rm amu}$) ions.
%The atomic data with the fast and slow ions are not present so that it is necessary to acquire the data for more accurate modeling.
%We applied the scaling laws to the calculation of the Fe line structures generated by impacts of various ions at $0.5$--$1000~{\rm MeV}/{\rm amu}$.
%This fact justifies the application of the scaling laws to the calculation of the Fe line structures generated by cosmic rays.
%%%

The intensity of each K$\alpha_1$L$^i$ (K$\alpha_2$L$^i$) sub-peak can be expressed as a function of $p_L^X$.
Under an assumption of independent L-shell electron ionization, a binomial distribution of
\begin{equation}
I_i= I_{tot}{8 \choose i}{p_L^X}^i(1-p_L^X)^{8-i}, \label{eq:Ii}
\end{equation}
where 
\begin{equation}
I_{tot} = \sum^8_i I_i, \label{eq:Itot}
\end{equation}
gives a good description of the relative intensities of the K$\alpha$L$^i$ peaks \citep{Kauffman1973}. 

According to \citet{Horvat2009}, intensities of the K$\alpha$L$^i$ peaks are changed depending on the phase of the target, whether it is gas or solid. 
Compared to the solid target case, the Fe K$\alpha$L$^i$ peaks become dominant with gaseous targets. 
\citet{Horvat2009} derived the scaling law of $p_L^X$ similar to the equation (\ref{eq:plx}) from atomic data of collisions between various ions at $\sim$ 10 MeV/amu and monoatomic Ar gas.
The scaling law for gas is the same as equation (\ref{eq:plx}), but with $a^{\prime}=0.856$, $b^{\prime}=2.94$, and $c^{\prime}=1.71$.

\subsection{Results}
\begin{figure*}
\begin{center}
 \includegraphics[width=18cm]{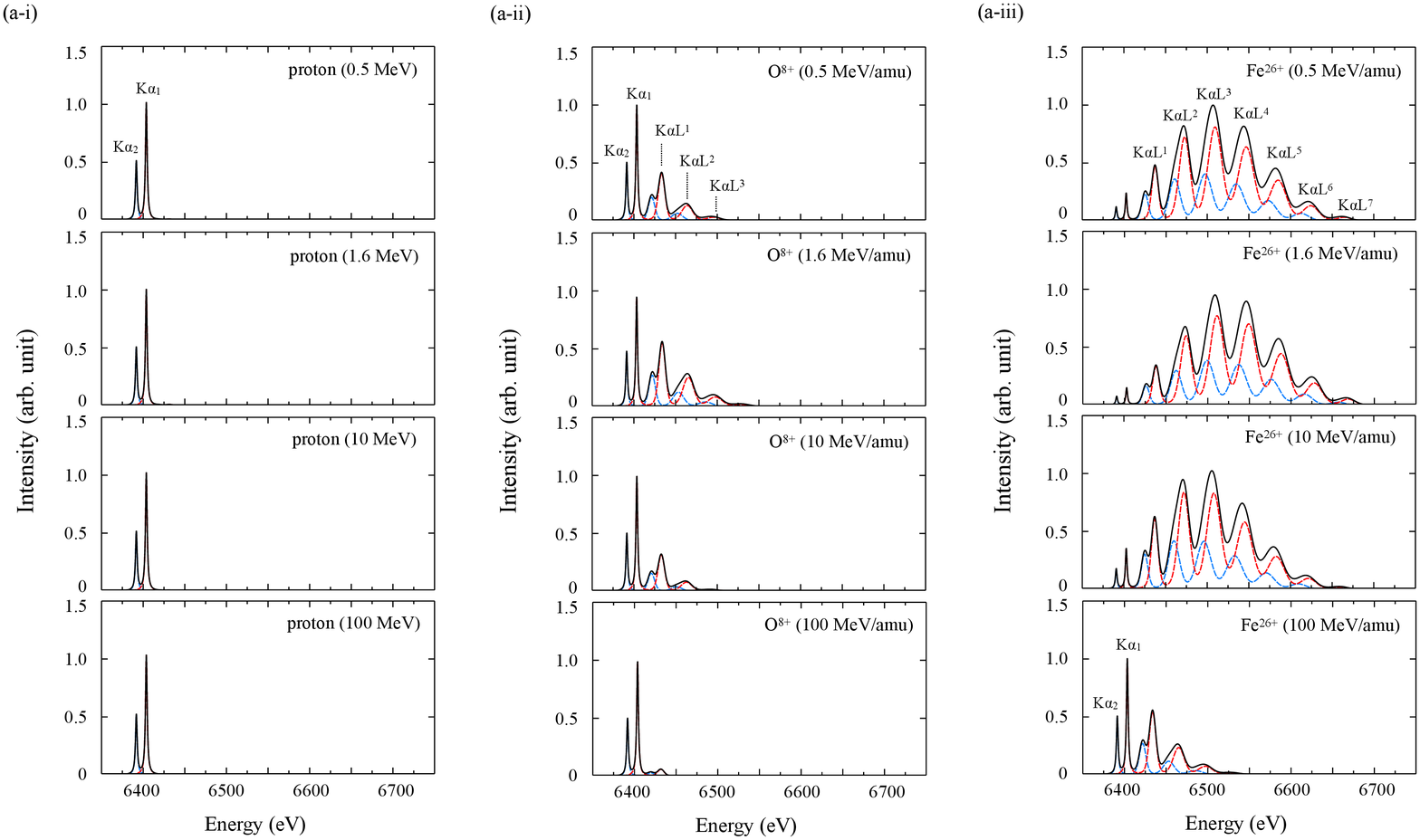} 
 \\
 \includegraphics[width=18cm]{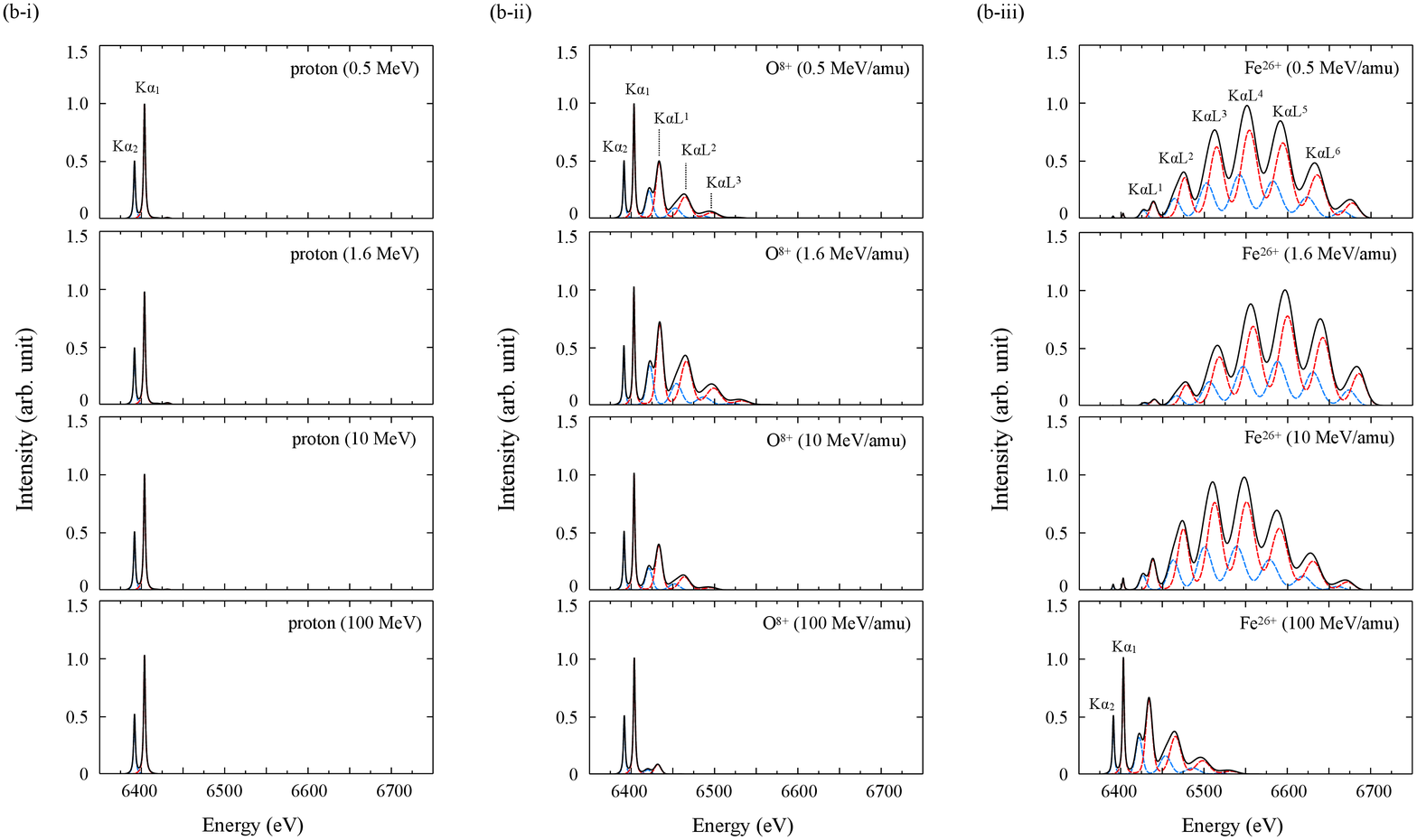} 
\vspace{0mm}
\end{center}
\caption{
Fe K$\alpha$ structures emitted from (a) solid and (b) gaseous Fe targets under bombardment by (i) protons, (ii) O$^{8+}$, and (iii) Fe$^{26+}$ ions.
The red and blue dashed curves denote the Fe K$\alpha_1$ and Fe K$\alpha_2$ structures, respectively.
}
\label{fig:Fig2}
\end{figure*}

In Figure~\ref{fig:Fig2}(a), we plot the models for the K$\alpha$ line emitted by solid Fe bombarded by protons 
and fully ionized O and Fe ions. 
Figure~\ref{fig:Fig2}(b) is the same but for gaseous Fe.  
The results clearly indicate that the intensities of the K$\alpha$L$^i$ peaks depends on the charge state and kinetic energy of the projectiles.
The larger charge the projectile has, the more significant the K$\alpha$L$^i$ peaks become. 
This is naturally explained by a stronger Coulomb field of heavy ions. 
Regardless of the projectile species, the Fe K$\alpha$L$^i$ peaks are the most significant when the projectile kinetic energy is $\sim 1.6~{\rm MeV}/{\rm amu}$.
Electrons generally are ejected most efficiently when the projectile ion velocity matches the average velocity of electrons in orbit of an atom \citep{Bohr1948}. 
Since the kinetic energy of $\sim 1.6~{\rm MeV}/{\rm amu}$ is translated into a projectile velocity close to that of Fe-L electrons, L-shell electrons are efficiently ionized 
around this energy. 

\begin{figure}
 \includegraphics[width=7.5cm]{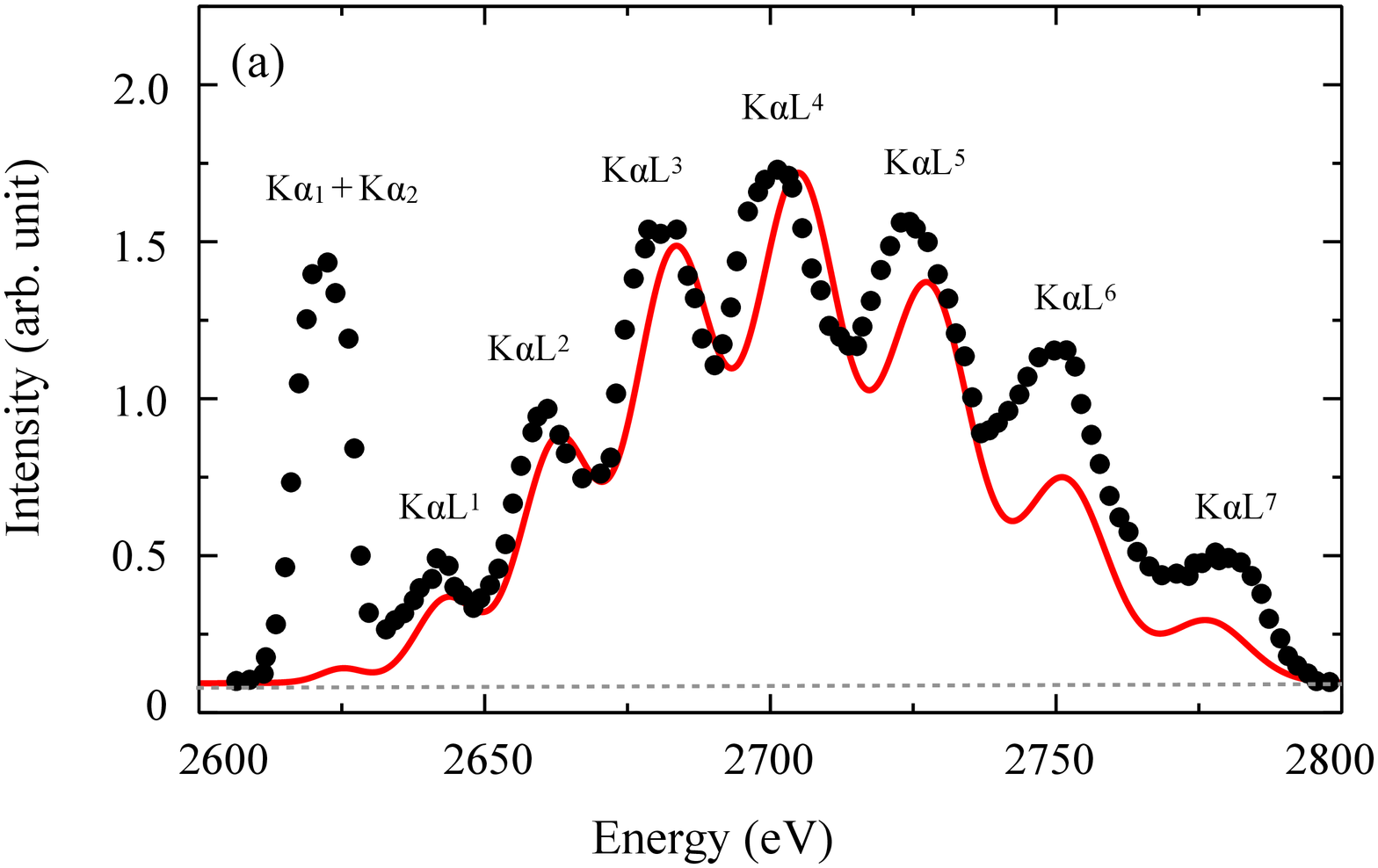} 
 \includegraphics[width=7.5cm]{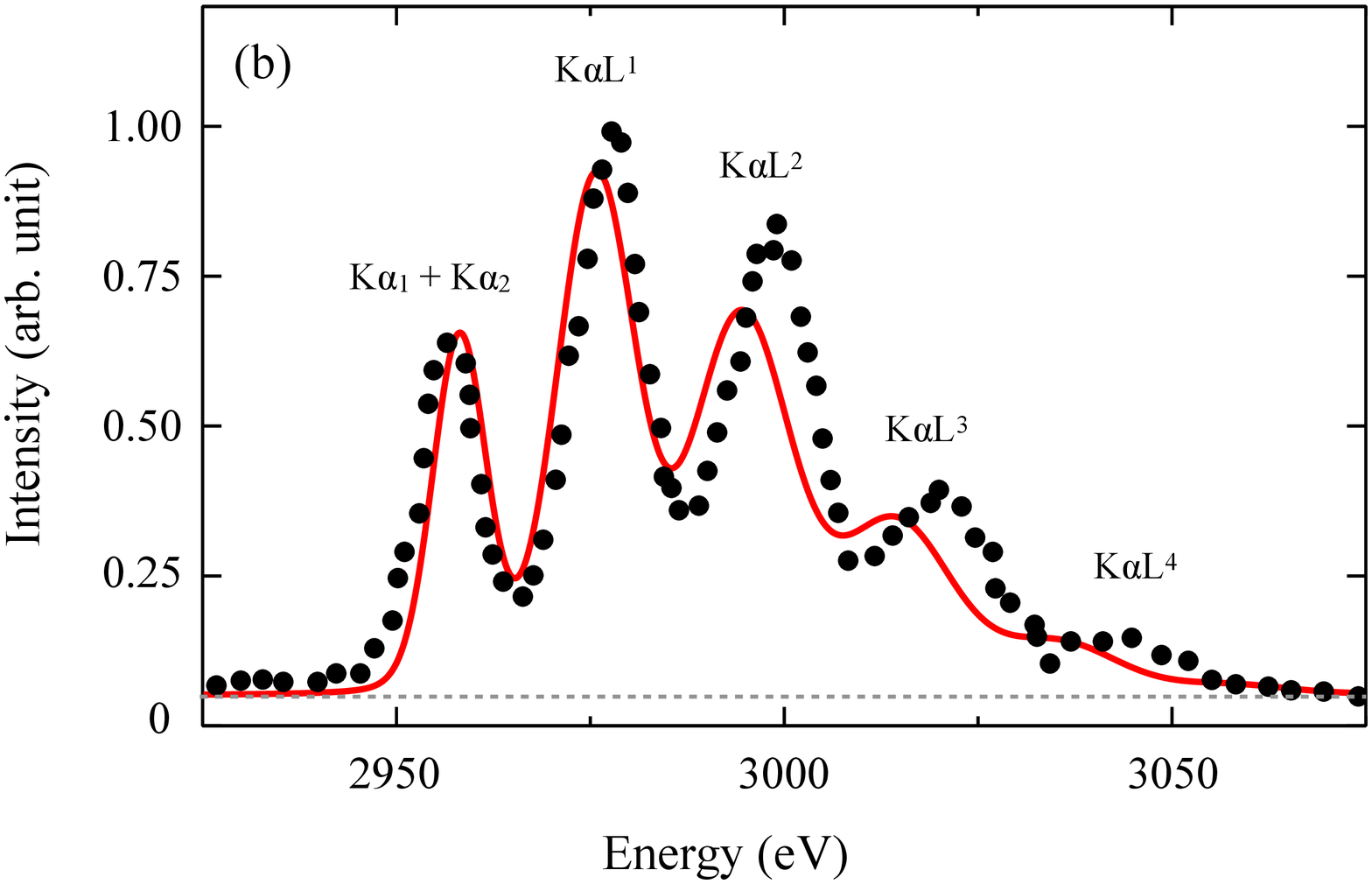} 
\caption{
(a) Cl K$\alpha$ spectrum from solid KCl targets under bombardment by $4~{\rm MeV}/{\rm amu}$ Kr$^{36+}$ ions \citep{Horvat2006} and 
(b) Ar K$\alpha$ spectrum from a gaseous Ar target under bombardment by $9.4~{\rm MeV}/{\rm amu}$ Ne$^{10+}$ ions \citep{Horvat2009} 
from experiments (the black dots), compared with our model (the red curves). 
The gray dashed curves represent the background.
}
\label{fig:Fig3}
\end{figure}

Figure \ref{fig:Fig3} shows a comparison between the computed models and experimental data by \citet{Horvat2006} and \citet{Horvat2009}.
In the experimental data, the K$\alpha$L$^i$ peaks ($i$ = 0--7) are generated by impacts of projectile ions, while the K$\alpha_1$ and K$\alpha_2$ lines in Figure~\ref{fig:Fig3}(a) are mainly induced by secondary electrons and X-rays.
Our model reproduces the overall K$\alpha$L$^i$ peaks strctures.
However, we found some discrepancies of the line center energies and intensities. 
The contradiction about the center energy could be explained by the inaccuracy of the equations (\ref{eq:delta_E}), (\ref{eq:plx}), and (\ref{eq:sigma}), 
which is estimated to be at a few $\%$ level by \citet{Horvat2006}. 
The discrepancy of the intensities could be brought by continuous changes of energy and charge state of projectile ions in thick ($\sim {\rm mg}~{\rm cm}^{-2}$) targets, which change the value of $p_L^X$.
Another possibility would be contribution of electron captures from the L-shell of the target atoms in L-shell vacancy production as \citet{Rymuza1989} and \citet{Kavcic2000} pointed out.
Those effects are not included in the present model.

%\section{Modeling Fe K$\alpha$ structures produced by cosmic-rays}
\section{Fe K$\alpha$ structures produced by CRs}

\subsection{Assumptions}

\begin{figure*}
\begin{tabular}{cccc}
\begin{minipage}[c]{0.5\hsize}
 \includegraphics[width=7.5cm]{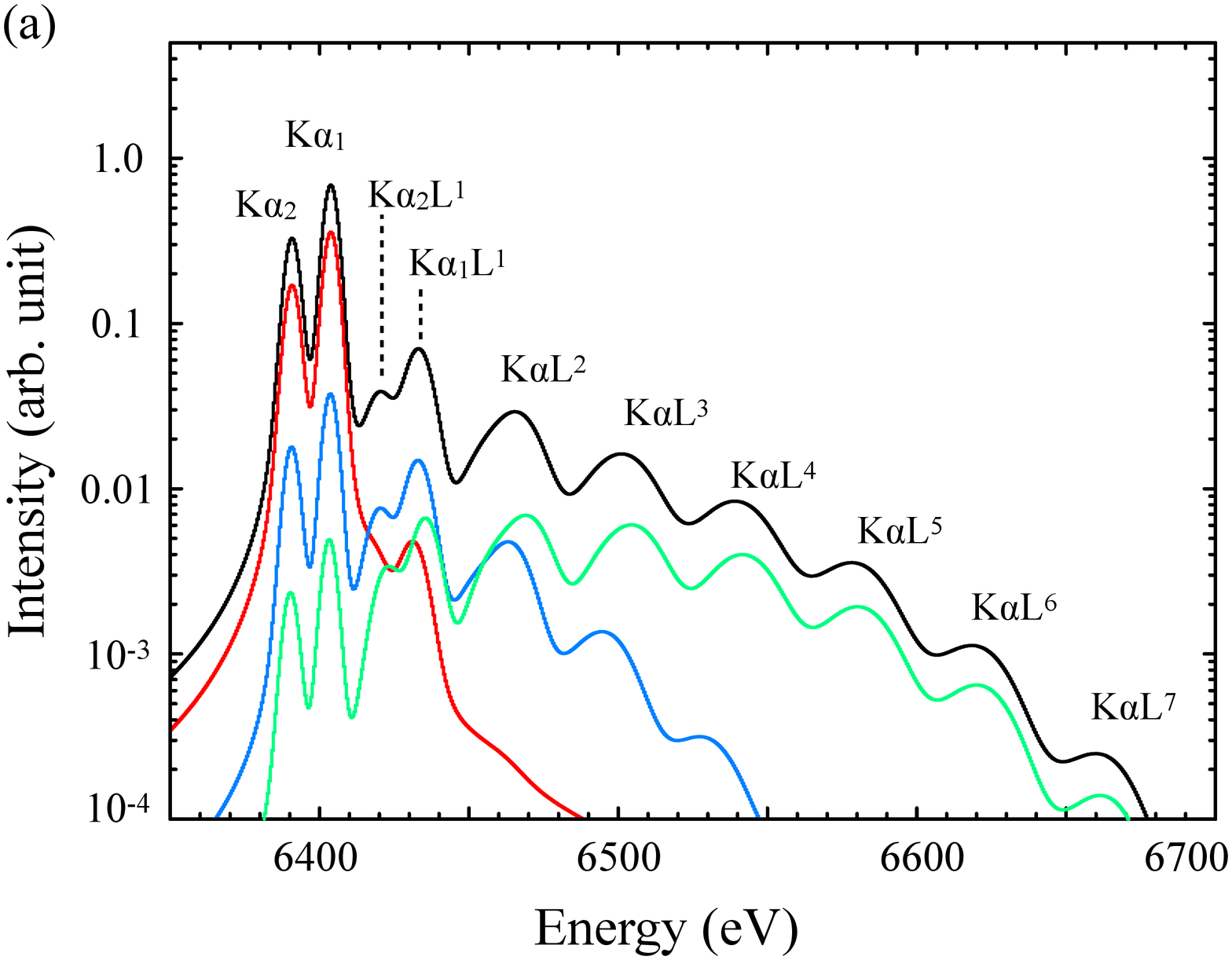} 
\end{minipage}
\begin{minipage}[c]{0.5\hsize}
 \includegraphics[width=7.5cm]{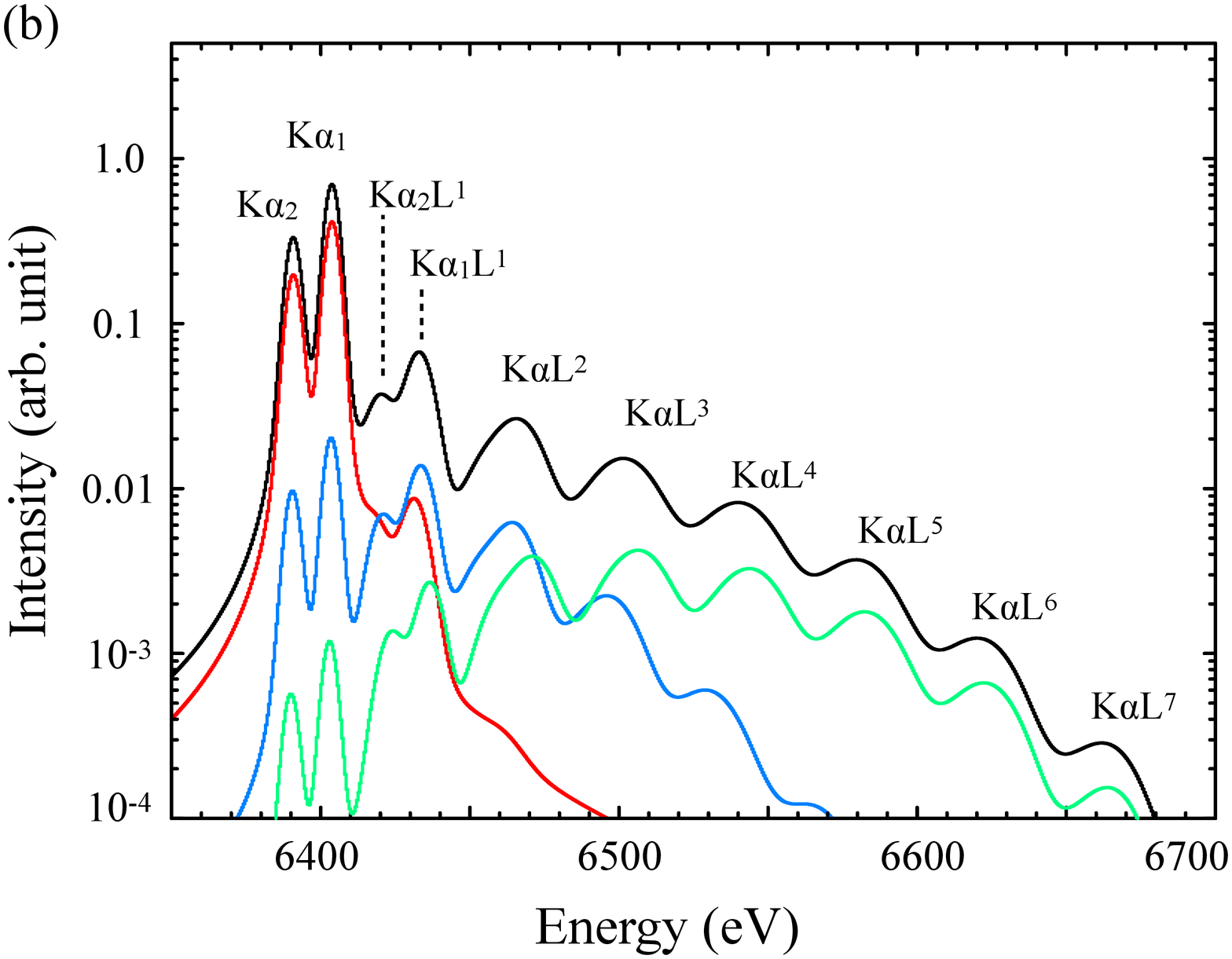} 
\end{minipage}
\end{tabular}
\caption{
Fe K$\alpha$ structures expected from Fe atoms bombarded by accelerated particles with spectral indices of (a) $s = 1$ and (b) $s = 2$ (black).
The red, blue, and green curves represent contributions from protons, O ions, and Fe ions, respectively.
}
\label{fig:Fig4}
\end{figure*}

Using our model, we compute Fe K$\alpha$ line spectra expected to be observed in SNRs. 
In SNR shocks, various ions would be accelerated at the same time, and thus the spectra would be superpositions 
of contributions from each ion species. 
Since the composition of CRs at acceleration sites is unknown, we here assume it is the same as that observed in the solar system by referring to \citet{Mewaldt1994}. 
We also assume that ions are fully ionized and that all ions follow a common spectrum. 
We take into account contributions from H, He, ..., Fe, Co, and Ni ions in an energy range of $0.5$--$1000~{\rm MeV}/{\rm amu}$. 

The flux of the Fe K$\alpha$L$^i$ peak $F_i$ by each ion species can be described as
\begin{equation}
F_i \propto n_{\rm gas}\int \sigma_{\rm KL^{\it i}}\,v_{\rm ion}\,\frac{dN_{\rm ion}}{dE}\,dE,
\end{equation}
where $n_{\rm gas}$, $\sigma_{\rm KL^{\it i}}$, $v_{\rm ion}$, and ${dN_{\rm ion}}/{dE}$ are the number density of Fe atoms in the dense gas, the production cross section of 
the Fe K$\alpha$L$^i$ peak, the velocity of the ions, and the differential spectrum of the ion, respectively.
We applied the scaling laws for solid targets assuming that Fe is mainly in dust grains. 
Based on equations (\ref{eq:Ii}) and (\ref{eq:Itot}), the production cross section $\sigma_{\rm KL^{\it i}}$ can be calculated as
\begin{equation}
\sigma_{\rm KL^{\it i}} = \sigma_{\rm p}{8 \choose i}{p_L^X}^i(1-p_L^X)^{8-i},~~\sigma_{\rm p}=\omega_{\rm K}\times\sigma_{\rm ion}, \label{eq3}
\end{equation}
where $\sigma_{\rm p}$, $\sigma_{\rm ion}$, and $\omega_{\rm K}$ are the production cross section of the Fe K$\alpha$ (=$\sum_i$ Fe K$\alpha$L$^i$ peaks) line, the K-shell ionization cross section, and the florescence yield (= 0.34) \citep{Krause1979}, respectively.
We computed $\sigma_{\rm ion}$ with the program by \citet{Batic2013} based on 
the ECPSSR (Energy-Loss Coulomb-Repulsion Perturbed-Stationary-State Relativistic) theory \citep{Brandt1981}.

\subsection{Results and future observations}
Figure \ref{fig:Fig4} presents the result, where the models are folded with the response of the X-ray micro-calorimeter Resolve \citep{Ishisaki2018} aboard XRISM \citep{Tashiro2018}.
Accelerated ions have a power-law spectrum $dN/dE\propto E^{-s}$ with $s = 1$ and $s = 2$ in the panels (a) and (b), respectively. 
The former is chosen, referring to the measurement of non-thermal bremsstrahlung in W49B by \citet{Tanaka2018}. 
The latter is the index expected in diffusive shock acceleration at a strong shock. 
In both cases, the K$\alpha$L$^i$ ($i  = 1$--$7$) peaks appear in the $6420$--$6700~{\rm eV}$ band.
The intensities of each K$\alpha$L$^i$ peak strongly reflect the composition of the emitting particles.
If particles accelerated in SNRs have a similar ion composition to CRs arriving at the solar system, 
the most significant peak is the K$\alpha$L$^1$ peak, which can be resolved into K$\alpha_1$L$^1$ and K$\alpha_2$L$^1$ with the Resolve.
The intensities of the two sub-peaks are $\sim 1/10$ of those of K$\alpha_1$ and K$\alpha_2$.
The Fe K$\alpha_1$L$^1$ and K$\alpha_2$L$^1$ sub-peaks are shifted by $\sim 30~{\rm eV}$ with respect to the Fe K$\alpha_1$ and K$\alpha_2$.

An $\sim {\rm eV}$ energy resolution by X-ray micro-calorimeters would be necessary to resolve the structures. 
In addition to XRISM, future missions such as Athena \citep{Barcons2017} and Super DIOS \citep{Ohashi2018} with X-ray micro-calorimeters will able to detect the line structures. 
Such studies will play a complimentary role to CR-ionization rate estimates based on H$_3^{+}$ absorption lines (e.g., \cite{Indrio2010}). 
We finally note that potential targets would not be limited to SNRs, considering the recent claims of contribution of CR-ionization to the neutral Fe K line 
in the Arches cluster region \citep{Tatischeff2012,Krivonos2017} and in the Galactic ridge \citep{Nobukawa2015}.

\begin{ack}
We thank T. Mukoyama, M. Pajek, B. Sulik, V. Tatischeff, and D. Banas for helpful discussion.
This work is supported by JSPS/MEXT Scientific Research Grant Numbers JP19J14025 (H.O.), JP16K06937 (M.I.), JP19H01936 (T.T.), 
JP19K03915 (H.U.), and JP15H02090 (T.G.T.). 
\end{ack}


\begin{thebibliography}{}

\bibitem[Ackermann et al.(2013)]{Ackermann2013} Ackermann, M., et al.\ 2013, Science, 339, 807  

\bibitem[Aharonian et al.(2007)]{Aharonian2007} Aharonian, F., et al.\ 2007, \aap, 464, 235  

\bibitem[Barcons et al.(2017)]{Barcons2017} Barcons, X., et al.\ 2017, Astronomische Nachrichten, 338, 153  

\bibitem[Bati{\v{c}} et al.(2013)]{Batic2013} Bati{\v{c}}, M., Pia, M.~G., \& Cipolla, S.~J.\ 2013, Computer Physics Communications, 184, 2232   

\bibitem[Bohr(1948)]{Bohr1948} Bohr, N., 1948, Det Kgl. Danske Videnskabernes Selskab. Math.-fys. Medd. XVIII, No.8.

\bibitem[Brandt \& Lapicki(1981)]{Brandt1981} Brandt, W., \& Lapicki, G.\ 1981, \pra, 23, 1717  

\bibitem[Burch et al.(1971)]{Burch1971} Burch, D., Richard, P., \& Blake, R.~L.\ 1971, \prl, 26, 1355  

\bibitem[Campbell \& Papp(2001)]{Campbell2001} Campbell, J.~L., \& Papp, T.\ 2001, Atomic Data and Nuclear Data Tables, 77, 1  

\bibitem[Dogiel et al.(2011)]{Dogiel2011} Dogiel, V., Chernyshov, D., Koyama, K., Nobukawa, M., \& Cheng, K.-S. 2011, PASJ, 63, 535

%\bibitem[Gaskin et al.(2019)]{Gaskin2019}	Gaskin, J.~A., et al.\ 2019, Journal of Astronomical Telescopes, Instruments, and Systems, 5, 021001  

\bibitem[Horvat et al.(2006)]{Horvat2006} Horvat, V., Watson, R.~L., \& Peng, Y.\ 2006, \pra, 74, 022718  

\bibitem[Horvat et al.(2009)]{Horvat2009} Horvat, V., Watson, R.~L., \& Peng, Y.\ 2009, \pra, 79, 012708  

\bibitem[Indriolo et al.(2010)]{Indrio2010} Indriolo, N., et al.\ 2010, \apj, 724, 1357

\bibitem[Ishisaki et al.(2018)]{Ishisaki2018} Ishisaki, Y., et al.\ 2018, Journal of Low Temperature Physics, 193, 991  

\bibitem[Kauffman et al.(1973)]{Kauffman1973}	Kauffman, R.~L., McGuire, J.~H., Richard, P., Moore, C.~F.\ 1973, \pra, 8, 1233  

\bibitem[Kav{\v{c}}i{\v{c}} et al.(2000)]{Kavcic2000} Kav{\v{c}}i{\v{c}}, M., et al.\ 2000, \pra, 61, 052711  

\bibitem[Krivonos et al.(2017)]{Krivonos2017} Krivonos, R., et al.\ 2017, \mnras, 468, 2822

\bibitem[Koyama et al.(1995)]{Koyama1995} Koyama, K., Petre, R., Gotthelf, E.~V., Hwang, U., Matsuura, M., Ozaki, M., \& Holt, S.~S. 1995, Nature, 378, 255

\bibitem[Krause(1979)]{Krause1979}	Krause, M.~O.\ 1979, Journal of Physical and Chemical Reference Data, 8, 307  

\bibitem[McGuire \& Richard(1973)]{McGuire1973} McGuire, J.~H., \& Richard, P.\ 1973, \pra, 8, 1374  

\bibitem[Makino et al.(2019)]{Makino2019} Makino, K., Fujita, Y., Nobukawa, K.~K., Matsumoto, H., \& Ohira, Y. 2019, PASJ, 71, 78

\bibitem[Mewaldt(1994)]{Mewaldt1994} Mewaldt, R.~A.\ 1994, Advances in Space Research, 14, 737  

\bibitem[Nobukawa et al.(2015)]{Nobukawa2015} Nobukawa, K.~K., et al.\ 2015, \apjl, 807, L10  

\bibitem[Nobukawa et al.(2018)]{Nobukawa2018} Nobukawa, K.~K., et al.\ 2018, \apj, 854, 87  

\bibitem[Nobukawa et al.(2019)]{Nobukawa2019} Nobukawa, K.~K., Hirayama, A., Shimaguchi, A., Fujita, Y., Nobukawa, M., \& Yamauchi, S. 2019, PASJ, 71, 115

\bibitem[Ohashi et al.(2018)]{Ohashi2018}	 Ohashi, T., et al. 2018, SPIE Proc., 10699, 1069928

\bibitem[Okon et al.(2018)]{Okon2018} Okon, H., Uchida, H., Tanaka, T., Matsumura, H., \& Tsuru, T.~G. 2018, PASJ, 70, 35

\bibitem[Rymuza et al.(1989)]{Rymuza1989} Rymuza, P., Sujkowski, Z., Carlen, M., Dousse, J.-C., Gasser, M., Kern, J., Perny, B., \& Rh{\'e}me, C. 1989,  Zeitschrift fur Physik D Atoms Molecules Clusters,  14, 37

\bibitem[Saji et al.(2018)]{Saji2018} Saji, S., Matsumoto, H., Nobukawa, M., Nobukawa, K.~K., Uchiyama, H., Yamauchi, S., \& Koyama, K. 2018, PASJ, 70, 23

%\bibitem[Savage \& Sembach(1996)]{Savage1996} Savage, B.~D., \& Sembach, K.~R.\ 1996, \araa, 34, 279

%\bibitem[Shevelko \& Tawara(2012)]{Shevelko2012} Shevelko, V., \& Tawara, H.\ 2012, Atomic Processes in Basic and Applied Physics

\bibitem[Sulik et al.(1987)]{Sulik1987} Sulik, B., K{\'a}d{\'a}r, I., Ricz, S., Varga, D., V{\'e}gh, J., Hock, G., \& Ber{\'e}nyi, D. 1987,  Nuclear Instruments and Methods in Physics Research B, 28, 509  

\bibitem[Tanaka et al.(2018)]{Tanaka2018} Tanaka, T., et al.\ 2018, \apjl, 866, L26  

\bibitem[Tashiro et al.(2018)]{Tashiro2018} Tashiro, M., et al. 2018, SPIE Proc., 10699, 1069922

\bibitem[Tatischeff et al.(2012)]{Tatischeff2012}	Tatischeff, V., Decourchelle, A., \& Maurin, G.\ 2012, \aap, 546, A88   

%\bibitem[Tolstikhina et al.(2018)]{Tolstikhina2018} Tolstikhina, I., Imai, M.,  Winckler, N., \& Shevelko, V.\ 2018, Basic Atomic Interactions of Accelerated Heavy Ions in Matter'

%\bibitem[Vriens(1967)]{Vriens1967} Vriens, L.\ 1967, Proceedings of the Physical Society, 90, 935 

\bibitem[Wang et al.(2012)]{Wang2012} Wang, X.-L., Dong, C.-Z., \& Su, M.-G.\ 2012, Nuclear Instruments and Methods in Physics Research B, 280, 93

%\bibliographystyle{plain}
%\bibliography{pasj} 

\end{thebibliography}
\end{document}